# Low spin-polarization in the heavy metal\ferromagnet structures detected through the domain wall motion by synchronized magnetic field and current

Xueying Zhang[1,2,3], Nicolas Vernier[3], Laurent Vila[4], Shaohua Yan[1,2], Zhiqiang Cao[1,2], Anni Cao[1], Zilu Wang[1], Wenlong Cai[1,2], Yang Liu[1], Huaiwen Yang[1,2], Dafiné Ravelosona[3], and Weisheng Zhao[1,2]*

[1]Fert Beijing Institute, BDBC, School of Microelectronics, Beihang University, Beijing 100191, China

[2]Beihang-Goertek Joint Microelectronics Institute, Qingdao Research Institute, Beihang University, Qingdao 266000, China

[3]Centre for Nanoscience and Nanotechnology, University Paris-Saclay, 91405 Orsay, France

[4]University Grenoble Alpes, CEA, CNRS, Grenoble INP, INAC, SPINTEC, Grenoble 38400, France

*weisheng.zhao@buaa.edu.cn

## Abstract

CoFeB is a very soft material, in which Domain Wall (DW) can be moved easily under a weak magnetic field. However, it is very difficult to move DWs in Ta\CoFeB\MgO nanowires with interfacial perpendicular magnetic anisotropy through a spin-polarized current, and this limits the perspectives of racetrack memory driven by the current-in-plane mechanism. To investigate this phenomenon, we performed experiments of DW velocity measurement by applying a magnetic field and a current simultaneously. Working in the precessional regime, we have been able to see a very important effect of the spin-polarized current, which allows evaluating the polarization rate of the charge carriers. An unexpected quite low spin polarization rate down to 0.26 have been obtained, which can explain the low efficiency of DW motion induced by the spin-polarized current. Possible reasons for this low rate are analyzed, such as the spin relaxation in the Ta layer.

In computing technology, memories with larger storage densities, higher speed and lower power consumption should be found. Magnetism provides interesting ways to get nonvolatile memories. One of the possibilities is the racetrack memory, in which the moving domain walls (DWs) are used to store and transfer the digital information [1,2]. However, up to now, this strategy has not succeeded, because moving DWs using only spin-polarized current has appeared to be more difficult than expected. Many materials have been studied [3]. But pinning effects are often too strong to make a reliable racetrack memory with them [4]. One of the promising materials was Ta\CoFeB\MgO, in which density of pinning sites seems very low, which enables movement of DWs using very low fields, as small as a few hundreds of μT [5–7]. However, although a quite good polarization rate of the charge carriers has been found in this material [8,9], very few authors have reported efficient movement in this material in zero fields using only spin-polarized current.

In addition, many physics problems in the heavy metal\ferromagnet structure [10–16], such as the spin transport properties, spin diffusion [17], must be understood. DW motions in Ta\CoFeB\MgO provides a good model to do this study.

To address these issues, we conduct the study on the DW motions induced by current and magnetic field in Ta\CoFeB\MgO narrow wires. First, we present several attempts to move domain walls using only spin-polarized current in nanowires made of Ta\CoFeB\MgO. These experiments resulted in a strange and non-convincing behavior, which lead us to try another type of experiment: moving the domain walls using both the spin-polarized current and magnetic field. This way, we could find a huge effect of the current. We could deduce from this an effective spin polarization ratio of the charge carriers. A possible explanation for this result is discussed.

The sample studied is a Ta(5nm)\Co$_{40}$Fe$_{40}$B$_{20}$(1nm)\MgO(2nm)\Ta(5nm) multilayers stack with perpendicular anisotropy, as shown in Fig. 1(a). It was annealed at 300 °C for two hours. Several properties of the samples have been experimentally characterized: saturation magnetization $M_S = 1.1 \times 10^6 A/m$, effective perpendicular anisotropy energy $K_{eff} = 2.2 \times 10^5 J/m^3$, the damping constant is α=0.013, and the DW width was estimated to be Δ= 10.7 nm [5], the DMI constant measured is less than $0.01mJ/m^2$ [18].

The sample is patterned by conventional electron beam lithography and ion beam etching into 50 μm long narrow wires connected with a 20μ×20μm square (nucleation pad). The width of the wire in these experiments are 1 μm or 1.5 μm. Golden electrodes are added on the magnetic structure for electrical characterization. Structure of the device and the electrical test configuration are shown in Fig. 1(b).

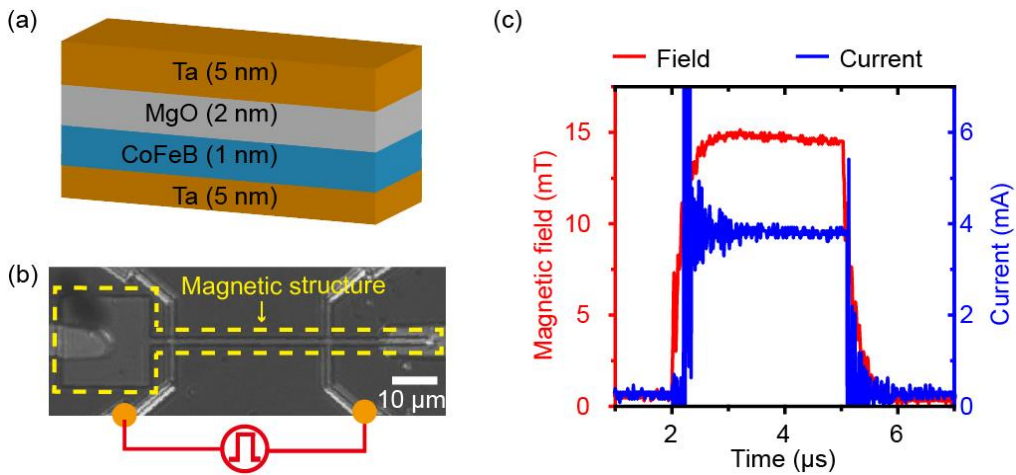

FIG. 1 (a) Stack structure of the studied sample. (b) Optical image of the tested device with a 1.5 μm wide wire and the configuration of electric measurement. (c) Example of the synchronized electric current pulse and magnetic field pulse applied on the tested sample. These pulses are supervised by an oscilloscope.

In the experiments, a homemade Kerr microscope with a resolution of 500 nm is used to observe the DW behavior in the device. A high voltage pulse generator is used to send current up to 15A during few μs in an 8-mm diameter coil. The small inductance of the mini coil, the large instantaneous output power and the ultrafast speed of the power supply allow obtaining large and fast field pulses. For example, magnetic field pulses with an amplitude of 130 mT and a rise time of 1.9 μs have been produced by a coil having 120 turns. The large amplitude is able to nucleate a DW in the micron-sized device and the short duration avoids the complete magnetic reversal of the structure, thus a DW can be obtained after the pulse. With a 20 turns coil, field pulses with a rise time of 220 ns have been obtained, so square form field pulses of several microseconds are obtained. These pulses are synchronized with the electric current pulse, as shown in Fig. 1 (c). A small delay between the field and the current is set, so that, the magnetic field has reached its plateau when the current started to flow in the wire.

In order to measure the DW velocity, first, after the DW nucleation, a Kerr image is acquired to locate the initial DW position. Then, a field (current, or both) pulse is applied and a second Kerr image is acquired to locate the final DW position. At last, the DW motion velocity is obtained by dividing the DW motion distance by the duration of the field (or current) pulse.

First, we have tried to induce the DW motion by the current alone. However, no regular DW motion has been observed even when the current density in the Ta\CoFeB layer has been increased to about $4.5 \times 10^{11} A/m^2$, as shown in Fig. 2(a)&(b). Although a little motion of DW in the direction of electron flow have been observed, these motions are not reproducible. When the current density further increases, the DW motion direction becomes stochastic: above a critical current density, DWs can go

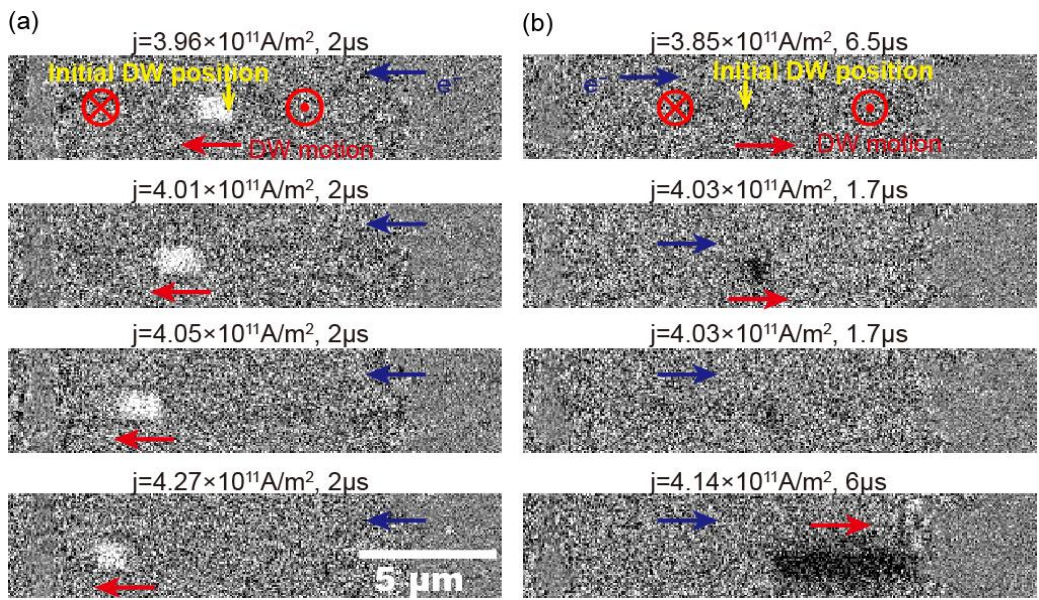

FIG. 2 (a) A series of DW motion to left and (b) to right, in a 1.0 μm wire driven by successive current pulses. Each picture is obtained by making the difference between a picture taken before the pulse and a picture taken after the pulse. So, the contrast shows the change of magnetization which has occurred during the pulse. Blue arrows indicate directions of electrons flow and red arrows indicate the DW motion directions.

as well in the direction of current or as opposite to this direction for the same pulse. The behavior of DWs is not reproducible and seems completely random. This is similar to the phenomena observed by S. Le Gall et al. [19].

In view of the difficulty of DW motion driven by current alone in this material, we begin to search for the regular and reproducible DW motions induced by the combined effects of the magnetic field and the electrical current. After DW nucleation, synchronized field pulses and current pulses are applied, all combinations of signs for the $B_\perp$ component and for the current have been checked. We found that, compared with DW motions induced by the magnetic field alone, the DW motion velocity changes obviously when the current is introduced. In high fields, for which the precessional regime is reached, the motion was accelerated in the direction of the electrons while it was slowed down when the DW motion is opposite to the direction of the electron flow, as shown in Fig. 3.

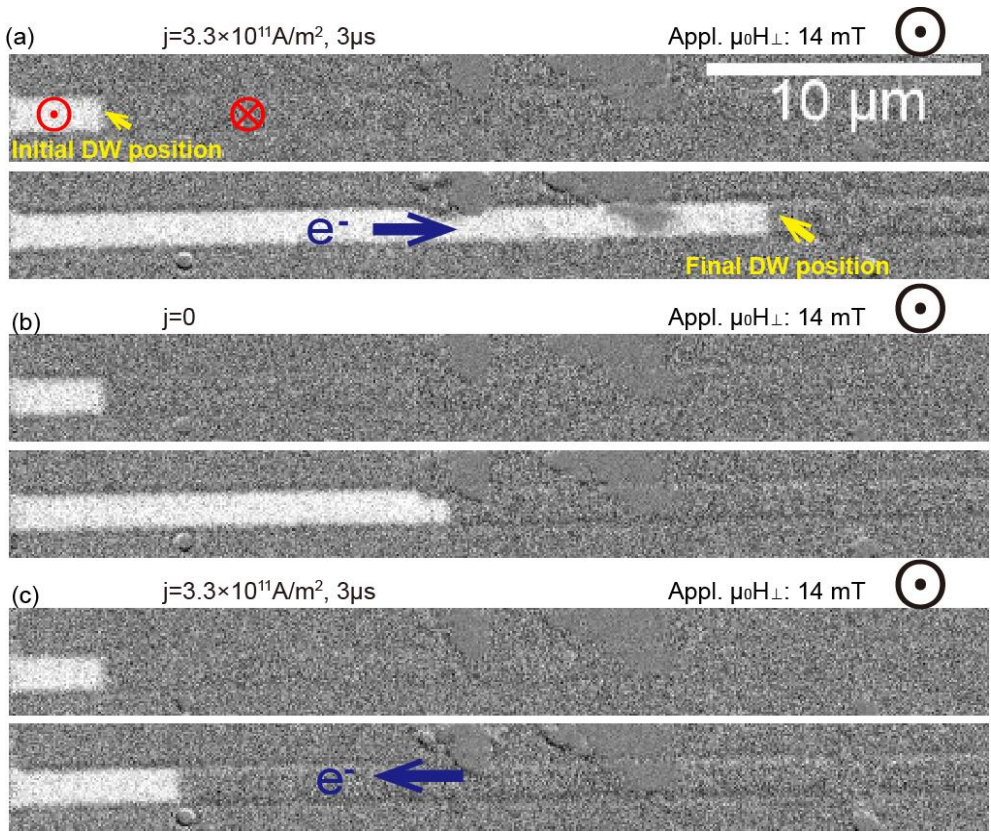

FIG.3 DW motion driven by synchronized current and field pulse. Pictures in each group give the DW positions before and after the pulse. Blue arrows indicate directions of electrons flow.

Velocities of DW motion with different current density and perpendicular field were measured systematically, as shown in Fig. 4(a). One can find that the DW velocity is obviously accelerated in the direction of electrons in the plateau (i.e., the DW motion when $\mu_0 H_\perp \geq 13$ mT).

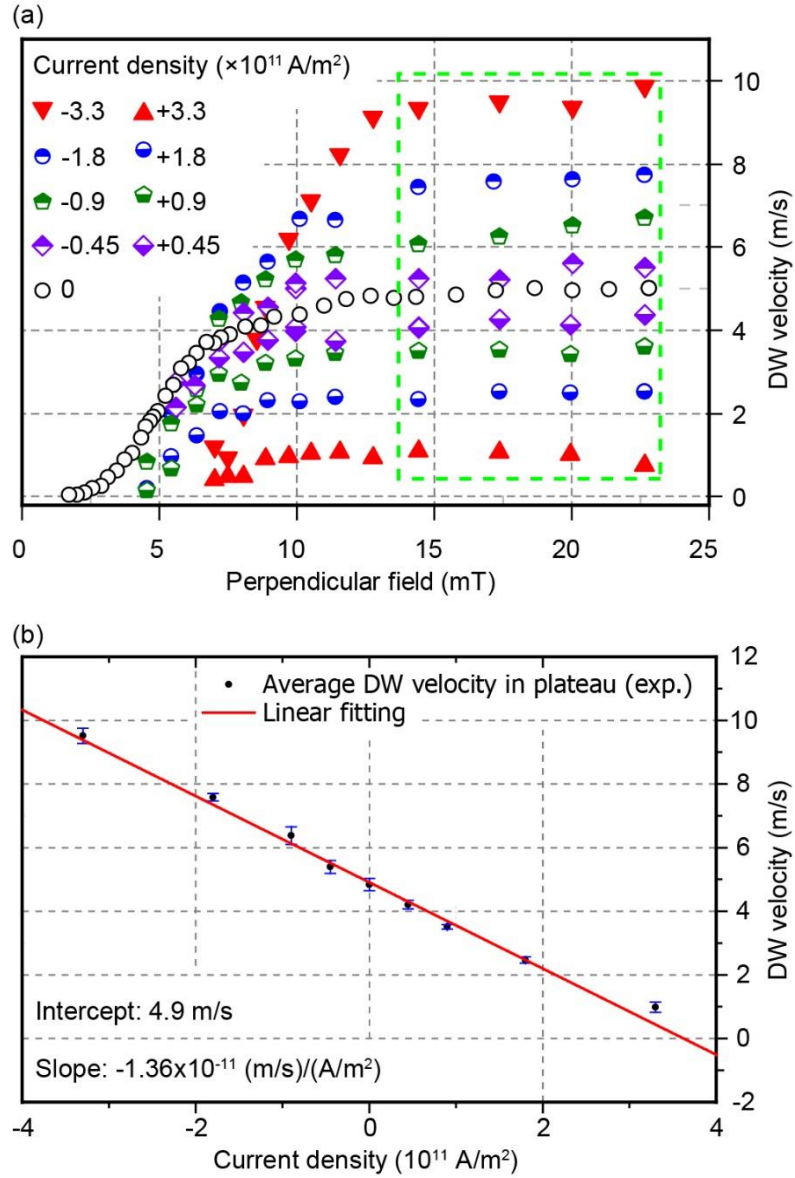

FIG.4 (a) DW motion velocities induced by different fields and current densities. Current flowing along the field-induced DW motion direction is defined as positive. Velocities in the plateau, boxed by the dashed green line, correspond the DW motion in the precessional regime. (b) Averaged velocities of DW motion shown in the plateau of (a) (enclosed in green). The error bar was taken as the standard deviation. The red line is a linear fitting on all these velocities. The current density in the CoFeB layer is estimated using $\rho_{CFB}$=170 $\mu\Omega\cdot$cm.

Theoretically, in a defect-free film, the field induced DW motion begins with the steady state regime at a low applied field, during which velocities of DW increase linearly with the magnitude of applied fields [20]. As the applied field is further increased, the Walker breakdown is reached, and the DW starts to move in the precessional regime. The Walker breakdown field can be calculated as $H_W = \alpha M_S/2$ [20]. In the sample studied here, this field is estimated to be as low as 0.8 mT due to the low damping. Experimentally, this theoritical behaviour has almost never been observed. Because of unavoidable defects in real samples, the DW motion is dominated by pinning effects for low driving fields. Typically, DW moves in the creep regime [21] or depinning transition regime [22] when the driving force is lower or approximately equals to the pining force. According to the field induced DW motion velocity measured on this device and the fitting with the creep law and depinning transition law, the creep regime and depinning transition regime end at a field of 4 mT and about 13 mT, respectively (see supplementary materials). Obviously, the plateau of DW velocity obtained in Fig. 4 (a) corresponds to the precessional regime while the Walker breakdown is masked by the pinning effects.

The dynamic of DW driven by a magnetic field along the easy axis and the STT can be described by the following one dimension (1 D) model [23,24],

$$\dot{\varphi} + \frac{\alpha}{\Delta}\dot{q} = \gamma_0 H_Z + \xi \frac{\mu_B P}{e\Delta M_s} j_{STT} \qquad (1)$$

$$-\alpha\dot{\varphi} + \frac{\dot{q}}{\Delta} = \frac{\gamma_0 H_K}{2} + \sin 2\varphi \frac{\mu_B P}{e\Delta M_s} j_{STT} \qquad (2)$$

Where $q$ is the position of the DW, $\varphi$ the angle between the magnetization at the center of the DW respect to the longitudinal direction of the wire (azimuth angle), $\gamma_0$ the gyromagnetic ratio, $\xi$ the non-adiabatic parameter, $H_K$ the shape anisotropy field of DW, $H_Z$ the applied field in the perpendicular direction, $\mu_B$ the Bohr magneton, e the electric charge, $P$ the spin polarization ratio and $j_{STT}$ the current density in the magnetic layer.

In the precessional regime, $\varphi$ alternates from 0 to $2\pi$, and the $\overline{\sin\varphi} = 0$ (the same for cos $\varphi$ and sin $2\varphi$). In this case, the Slonczewski torque from the eventual spin Hall current arising from the sub-Ta layer is zero. Our observation is consistent with the fact that the DMI is very weak in this sample; if not, the DMI will fix the magnetization of the DW and no precession will occur at relatively low fields [25].

By assuming $\overline{\sin 2\varphi} = 0$, we can get,

$$v_{STT+B} = v_{STT} + v_B \qquad (3)$$

With

$$v_{STT} = \frac{1+\alpha\xi}{1+\alpha^2} \cdot \frac{\mu_B P}{eM_s} j_{STT} \qquad (4)$$

In view that $\alpha \ll 1$, $v_{STT}$ can be approximated by the spin drift velocity [24],

$$v_{STT} = \frac{\mu_B P}{eM_s} j_{STT} \qquad (5)$$

Now, let us return to the experimental results shown in Fig.4(a). In the precessional mode, DW velocities remains stable in this range, we did an average on the velocities (enclosed in green dashed line in Fig. 4(a)) for each current and plotted the mean velocities as a function of the applied current, as shown in Fig. 4(b). A very good linear relationship between the DW velocity and the magnitude of the applied current was obtained, consistent with the above analysis (Eq. 3 - 5). After a linear fitting, an intercept of 4.9 m/s and a slope $k = -1.36 \times 10^{-11} \, (m/s)/(A/m^2)$ are obtained. Note that in this plot, we assume that the current density in each conductive layer is proportional to its conductivity, thus, the current density in the CoFeB layer can be estimated with $j_{STT} = \frac{U}{l\rho_{CFB}}$, where U is the potential difference between the two terminals of the measured wire, $l$ is the length of the wire and $\rho_{CFB}$ is the resistivity of CoFeB. According to our experiments and the published litteratures [3,26,27] (see supplementary materials), $\rho_{CFB} = 170 \, \mu\Omega \cdot cm$ is used.

In fact, the relationship described in Eq. (3) has been observed in some experiment performed on materials with the in-plane anisotropy such as the permalloy [1,28,29] or on the [CoNi]n superlattice with PMA [19]. However, to our knowledge, it is the first time that this relationship is observed in the heavy metal\ferromagnet\insulator multilayers. In the linear relationship described in Fig.4, the intercept corresponds to the DW velocity induced by the magnetic field alone and the slope represents the contribution from the current-induced motion. Based on this slope and the relationship described in Eq. (5), we can get the effective polarization of CoFeB in these experiments: $P$= 0.26.

The polarization ratio obtained by measuring the in-plane current-induced DW motion is much lower than the intrinsic spin polarization ratio of CoFeB measured through the Point-Contact Andreev Reflection (65%) [8], through the superconducting tunneling spectroscopy (53%) [9,30,31] in the heavy metal\CoFeB\Oxide layer structure or through spin absorption experiment on $Co_{60}Fe_{40}$ in lateral spin valves (48%) [17].

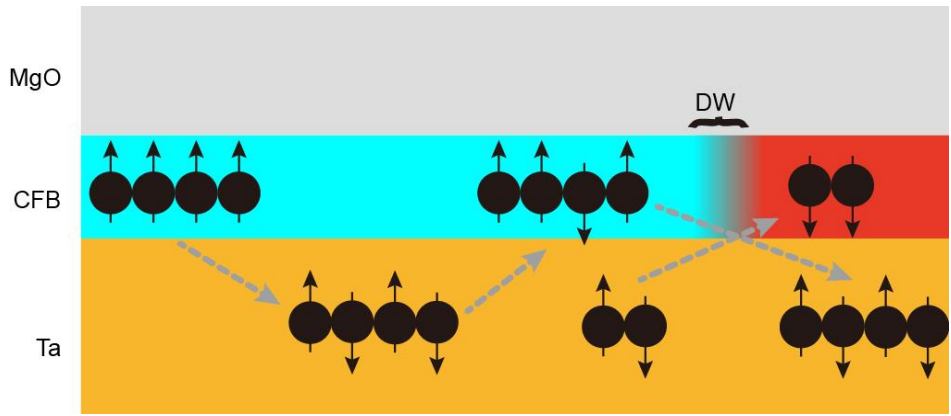

FIG.5 Sketch of the electrons transport in the Ta/CoFeB/MgO multilayers structure.

The low polarization measured through the STT induced DW motion in Ta\CoFeB\MgO structure may be caused by the following reasons. First, because of scattering, the conductive electrons interpenetrate between the Ta layer and CoFeB layer, as illustrated in Fig. 5. The spin relaxation in Ta layer, including the diffusion of non-spin-polarized electrons from the Ta layer into the FM layer, and the absorption of spin by Ta [17], and its interface, may reduce the effective spin polarization of CoFeB. Second, the spin injection into the CoFeB layer, or spin accumulation at the Ta\CoFeB interface due to spin Hall and Rashba effects may reduce the spin polarization of CoFeB.

Note that we have found that it was difficult to observe the DW motion driven by electric current alone in this structure. The low effective spin polarization for current flowing in the plane of Ta\CoFeB may be one of the reasons to explain this difficulty. These explanations are consistent with a recent result: it has been found that DWs could be moved quite easily in the same kind of sample if the thickness of Ta is reduced [32,33]. Indeed, in this case, diffusion in the Ta layer is reduced and we can expect a better spin polarization rate of the charge carriers making the DWs motion easier.

In conclusion, the DW motion in Ta\CoFeB\MgO narrow wire induced by the combined effect of the magnetic field and the electrical current has been observed and measured with Kerr microscopy. In the precessional regime, the DW velocity is found to be the linear addition of the field-induced velocity and the spin-polarized current drift velocity. The spin polarization of the CoFeB in this structure has been extracted based on the dependence of the DW velocity on the current density, which is found to be much lower than the polarization in the case where current flows perpendicularly to the plane. This decay of spin polarization may be explained by the interpenetration of electrons between the heavy metal and ferromagnet layers and the strong spin-orbit coupling in the heavy metal layer. This result can explain the long-standing difficulty of the current induced DW motion in similar structures.

**Acknowledgments**

The authors gratefully acknowledge the National Natural Science Foundation of China (Grant No. 61627813 and 61571023), Programme of Introducing Talents of Discipline to Universities (Grant No. B16001), the National Key Technology Program of China 2017ZX01032101 and the China Scholarship Council.

# Low spin-polarization in the heavy metal\ferromagnet structures detected through the domain wall motion by synchronized magnetic field and current

Xueying Zhang[1,2,3], Nicolas Vernier[3], Laurent Vila[4], Shaohua Yan[1,2], Zhiqiang Cao[1,2], Anni Cao[1,2], Zilu Wang[1], Wenlong Cai[1,2], Yang Liu[1], Huaiwen Yang[1,2], Dafiné Ravelosona[3], and Weisheng Zhao[1,2]*

*1Fert Beijing Institute, BDBC, School of Microelectronics, Beihang University, Beijing 100191, China*

*2Beihang-Goertek Joint Microelectronics Institute, Qingdao Research Institute, Beihang University, Qingdao 266000, China*

*3Centre for Nanoscience and Nanotechnology, University Paris-Saclay, 91405 Orsay, France*

*4University Grenoble Alpes, CEA, CNRS, Grenoble INP, INAC, SPINTEC, Grenoble 38400, France*

*weisheng.zhao@buaa.edu.cn*

## *Supplementary materials*

### 1. DW motion induced by perpendicular magnetic fields

Velocities of DW motion induced by perpendicular magnetic fields in a 1.5 μm wide wire are measured, as shown in Fig. S1. As the magnitude of the filed increases, the DW motion undergoes the creep regime for field smaller than 4 mT, the depinning transition regime for field between 4 mT and 13 mT and the precessional regime for field larger than 13 mT.

First, for the creep motion, the experimentally measured results are fitted with the creep law [1,2],

$$v = v_0 \exp\left[-\left(\frac{U_C}{k_B T}\right)\left(\frac{H_{p\_intr}}{H}\right)^{\mu}\right] \qquad (S-1)$$

Where $v$ is the DW velocity and $H$ is the applied magnetic field along the easy axis. $U_C$ is a parameter characterizing the strength the energy barriers due to defects in the material, $k_B$ is the Boltzmann constant, T is the temperature, $v_0$ is constant related to the properties of the material with a dimension of velocity, and μ=1/4, is a constant. $H_{p\_intr}$ is called the intrinsic pinning field of a material. The theory of creep regime applies from zeros field until $H_{p\_intr}$, i.e., the creep law ends when the applied field reaches $H_{p\_intr}$. Usually, this critical field can be seen as an indicator of the strength of the pinning effects of the film.

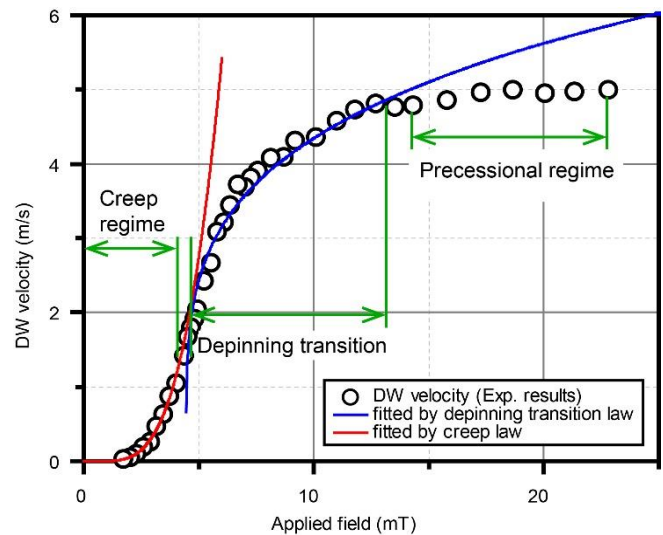

FIG. S1 Experimentally measured velocities of field induced DW motion in a 1.5 μm wire and the fitting results with the creep law and depinning transition law (in red and blue line, respectively).

Based on the creep law, the measured DW velocities $v$ vs. the applied field $\mu_0 H$ has been fitted with the following formula,

$$v = v_0 \exp\left[A \cdot (\mu_0 H)^{-1/4}\right] \tag{S-2}$$

Where $A = -\left(\frac{U_c}{k_B T}\right)\left(\mu_0 H_{p\_intr}\right)^{1/4}$ and $v_0$ are set as the unknown parameters during the fitting. As shown in Fig. S1, the creep law fits very well with measured results for weak fields and we get $v_0 = 1.2 \times 10^7 \ m/s$ and $A = 22.89$ (in $mT^{1/4}$). DW velocities leave the creep law after 4 mT, meaning that the intrinsic pinning field of the studied sample is about 4 mT, which is a relatively low value.

After the creep motion, the experimental result has been fitted with the depinning transition law, expressed as [3],

$$v = v_d \left(\frac{H - H_{p\_intr}}{H_{p\_intr}}\right)^{\beta} \tag{S-3}$$

Here, $v_d$ and $H_{p\_intr}$ is set as unknown parameters and $\beta$ is set as a constant 0.25, as experimentally demonstrated by R. Diaz Pardo et al [3]. As shown in Fig. S1, the measured DW velocities can fit very well with depinning transition law for fields in the range from 4.5 mT to 13 mT. We extracted the following parameters, $v_d = 4.1 \ m/s$ and $\mu_0 H_{p\_intr} = 4.5 \ mT$, which get a relatively good agreement with the intrinsic pinning field obtained by observing the end of the creep law.

For magnetic fields beyond 13 mT, the DW motion velocities reach a plateau. In fact, the Walker breakdown field $\mu_0 H_W$ is estimated to be 0.8 mT in this sample using the formula $H_W = \alpha M_S/2$ [4] and this breakdown is masked by the pinning effects. The plateau we obtained here is in the precessional mode.

## 2. Estimation of the resistivity of CoFeB thin layer in the studied sample

According to the published results, the resistivity of CoFeB $\rho_{CFB}$ thin films measured in similar structures varies from 160 to 330 $\mu\Omega\cdot$cm, as listed in Table S1. In view of this large variation and the key importance of $\rho_{CFB}$ for the extraction of the polarization of CoFeB, we have performed some experiments to estimate the CoFeB thin film in our sample.

Table S1 Resistivity of CoFeB in measured similar structures according to published results

| Resistivity of CFB (μΩ·cm) | Resistivity of Ta (μΩ·cm) | Composition and structure (thickness in nm) | Thickness of CFB (nm) | Annealing condition | Measurement method | Reference |
|---|---|---|---|---|---|---|
| 160 | 189 | Sub/TaN(t)/CoFeB(1)/MgO(2)/Ta(1) | 1 | 300°C, 1 h | R vs. thickness of TaN in wire | J. Torrejon et al. (2014), [8] |
| 170 | 190 | Si/Co40Fe40B20/Ta(8) | 4 | unknown | R vs. thickness of Ta in wire | Luqiao Liu et al. (2012), [7] |
| 330 | -- | W(2-7)/Co32Fe48B20/MgO | 0.8-1.4 | 250°C, 30 min | R vs. thickness of W in wire (measured by TR-MOKE) | S. Cho et al. (2015) [9] |
| 140 | -- | Co60Fe20B20 | 100 | 350°C,1h | four-point sheet resistance meter | Y.-T. Chen et al. (2013) [10] |
| 250 | -- | Co40Fe40B20 | 10 | No | four- point sheet resistance meter | Y.-T. Chen et al. (2012) [11] |
| 114 | -- | Co40Fe40B20 | bulk | Amorphous | | W. Kettler et al. (1982) [12] |

As shown in S2 (b) & (c), 2cm×2cm films with two different layer structures are deposited using sputtering: Substrate\MgO(2)\CoFeB(t)\Ta(5) and Substrate\Ta(3)\CoFeB(t)\MgO(2)\Ta(2) with thickness in nm and t varying from 0 to 3. All these samples are annealed at 300°C for 2 hours and the composition are all $Co_{40}Fe_{40}B_{20}$. The square conductance is measured using the four probes method [5]. The measuring steps are: first, four-point probes were connected in the center of a 2cm ×2cm sample. Four probes are connected to the samples. The four probes form a square in the center of the sample and the size of the square is much smaller than the sample. Second, a current $I_{i,j}$ is applied via two of the neighboring probes and the voltage $V_{k,l}$is measured via the other two probes, where $i$, $j$, $k$, $l$ are the index of the probes. A resistance was obtained by $R_{ij,kl} = V_{k,l}/I_{i,j}$. Then, this measurement was repeated by changing the probes applying the current. At last, the square resistivity of the multi-layer stack can be obtained with the following formula [5],

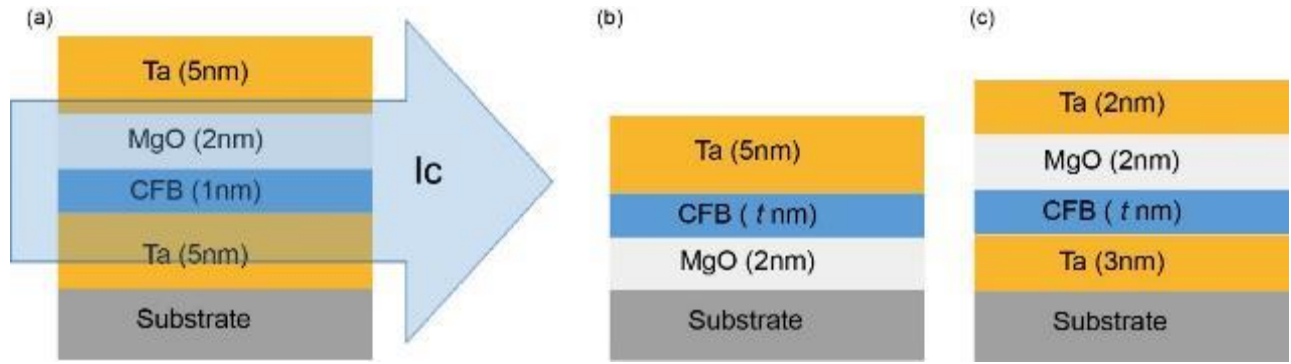

FIG. S2 (a) The multilayers structure of the sample studied; (b)&(c) Multilayers structures with thinner Ta layers and varying thickness of CoFeB used for the measurement of the resistivity of CoFeB.

$$R_S = \frac{\pi}{2\ln 2}\left(R_{12,34} + R_{23,41} + R_{34,12} + R_{41,23}\right) \qquad \text{(S-4)}$$

The square conductance can be defined as $G_S = 1/R_S$. If the conductance of the multilayers stack can be seen as the accumulation of the conductance of the Ta layers and the CoFeB layer,

$$G_S = \sigma_{CFB} t_{CFB} + \sigma_{Ta} t_{Ta} \qquad \text{(S-5)}$$

where $\sigma_{CFB}$ and $\sigma_{Ta}$ are the conductivity of CoFeB and Ta, $t_{CFB}$ and $t_{Ta}$ are the thickness of CoFeB and Ta. If the thickness dependence of conductivity (Fuchs-Sondheimer model) [6] is neglected, the conductance of the CoFeB can be extracted from the variation of Gs vs. the thickness of CoFeB, as plotted in Fig S3.

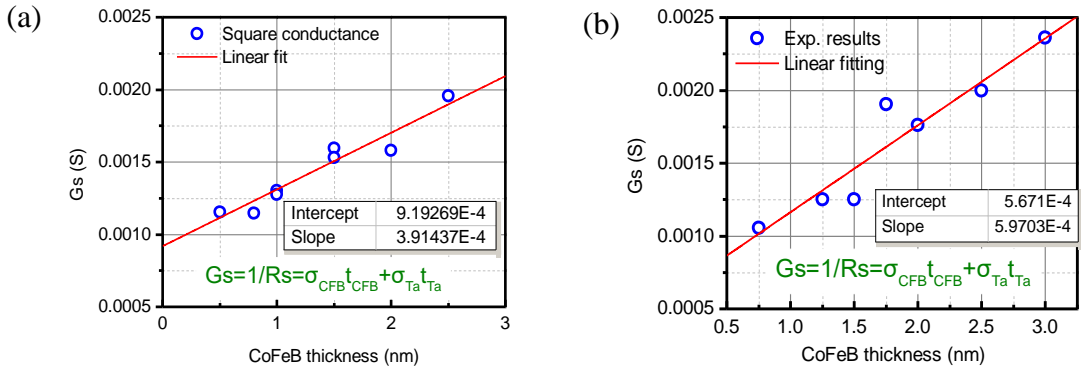

FIG. S3 Square conductance of the film vs. the thickness of CoFeB in (a) the MgO(2)\CoFeB(t)\Ta(5) structure and (b) in the \Ta(3)\CoFeB(t)\MgO(2)\Ta(2) structure. Note that the measurement of the square conductance on the sample with 0nm CoFeB (a) and the sample with 0 and 0.5nm CoFeB (b) was failed because of the high contact resistance.

According to the measurements on the Sub\MgO2)\CoFeB(t)\Ta(5) structure, the resistivity of the CoFeB is extracted to be 256 μΩ·cm; according to the measurements on the Sub\Ta(3)\CoFeB(t)\MgO(2)\Ta(5) structure, the resistivity of CoFeB is extracted to be 168 μΩ·cm. These results are close to the results measured by L. Liu et al. (2012), [7], J. Torrejon et al. (2014), [8]. Therefore, in this article, we adopt the results of Liu et al. (2012) [7], i.e., $\rho_{CFB} = 170\ \mu\Omega \cdot cm$.